\documentclass[a4paper,11pt]{article}
\pdfoutput=1

\usepackage{a4}
\usepackage{jheppub}
\usepackage{graphics}
\usepackage{amssymb,amsmath}
\usepackage{color}
\usepackage{wasysym}
\usepackage[tight]{subfigure}
\usepackage{booktabs}

\newcommand{\beq}{\begin{equation}}
\newcommand{\eeq}{\end{equation}}
\newcommand{\beqn}{\begin{eqnarray}}
\newcommand{\eeqn}{\end{eqnarray}}

\newcommand{\sstop}{\tilde{t}_1}

\newcommand{\chap}[1]{\tilde{\chi}^+_{#1}}
\newcommand{\cham}[1]{\tilde{\chi}^-_{#1}}
\newcommand{\chapm}[1]{\tilde{\chi}^\pm_{#1}}
\newcommand{\mneu}[1]{m_{\tilde{\chi}^0_{#1}}}
\newcommand{\mcha}[1]{m_{\tilde{\chi}^\pm_{#1}}}
\newcommand{\mstop}[1]{m_{\tilde{t}_{#1}}}
\newcommand{\neu}[1]{\tilde{\chi}^0_{#1}}
\newcommand{\cha}[1]{\tilde{\chi}^\pm_{#1}}

\newcommand{\gev}{\ \mathrm{GeV}}
\newcommand{\tev}{\ \mathrm{TeV}}
\newcommand{\ifb}{\ \mathrm{fb}^{-1}}

\newcommand{\met}{$E_T^\mathrm{miss}$}
\newcommand{\mathmet}{E_T^\mathrm{miss}}

\newcommand{\invfb}{fb${}^{-1}$}

\newcommand{\ATLAS}{A\textsc{tlas}}
\newcommand{\CMS}{C\textsc{ms}}
\newcommand{\Delphes}{D\textsc{elphes}}
\newcommand{\CheckMATE}{Check\textsc{mate}}
\newcommand{\Checkmate}{Check\textsc{mate}}
\newcommand{\ATOM}{A\textsc{tom}}

\newcommand{\LHC}{L\textsc{hc}}
\newcommand{\Lhc}{L\textsc{hc}}
\newcommand{\NLLFast}{N\textsc{ll-fast}}
\newcommand{\Herwig}{H\textsc{erwig++}}
\newcommand{\Rivet}{R\textsc{ivet}}
\newcommand{\Prospino}{P\textsc{rospino-2}}
\newcommand{\Fastjet}{F\textsc{astJet}}

\preprint{{\tiny KCL-PH-TH/2014-26, LCTS/2014-24, IFT-UAM/CSIC-14-050}}
\title{`Stop' that ambulance! New physics at the LHC?}
\author[a]{Jong Soo Kim,}
\author[a]{Krzysztof Rolbiecki,}
\author[b]{Kazuki Sakurai,}
\author[c]{and Jamie Tattersall}
\affiliation[a]{
Instituto de F\'{\i}sica Te\'{o}rica, IFT-UAM/CSIC,\\
C/ Nicol\'{a}s Cabrera, 13-15, Cantoblanco, 28049 Madrid, Spain}
\affiliation[b]{King's College London, \\ Department of Physics, London WC2R 2LS, UK}
\affiliation[c]{University of Heidelberg, Institut f\"ur Theoretische Physik,\\ Philosophenweg 16, D-69120 
Heidelberg, Germany}
\emailAdd{jong.kim@csic.es}
\emailAdd{krzysztof.rolbiecki@desy.de}
\emailAdd{kazuki.sakurai@kcl.ac.uk}
\emailAdd{tattersall@thphys.uni-heidelberg.de}
\abstract{
A number of LHC searches now display intriguing excesses. Most prominently, the measurement of the $W^+W^-$ cross-section
has been consistently $\sim 20\%$ higher than the theoretical prediction across both \ATLAS\, and \CMS\, for both 7 and 8~TeV runs. More recently, supersymmetric searches
for final states containing two or three leptons have also seen more events than predicted in certain signal regions. We show 
that a supersymmetric model containing a light stop, winos and binos can consistently match the data. We perform a fit to all measurements
and searches that may be sensitive to our model and find a reduction in the log-likelihood of 15.4 compared to the Standard Model 
which corresponds to 3.5-$\sigma$ once the extra degrees of freedom in the fit are considered. 

}

\begin{document}
\maketitle
\flushbottom


\section{Introduction\label{intro}}
After two years of running, the \ATLAS\, and \CMS\, experiments at the Large Hadron Collider (\Lhc) have performed a large number of 
Standard Model (SM) measurements and searches of beyond-Standard-Model (BSM) physics. Despite an overall agreement 
between experimental results and SM predictions, several 2-$\sigma$ excesses can be spotted in some key areas of the data. 
While one would normally expect some of these measurements to fluctuate around the central values, the picture becomes 
more intriguing when the excesses point to similar kinematic region and final states. In this paper we re-analyse \LHC\ 
measurements to test a hypothesis that the overall agreement between the data and theory predictions can be improved by 
including the production of light supersymmetric \cite{Martin:1997ns} top quark partners~\cite{Rolbiecki:2013fia}. This task is performed using 
two automated computer codes aimed at confronting new physics models with the \LHC\, data, \ATOM~\cite{ATOM} 
and \CheckMATE~\cite{Drees:2013wra}.

A starting point of our analysis are the $W^+W^-$ cross section measurements performed by \ATLAS\, 
and \CMS~\cite{ATLAS:2012mec,Chatrchyan:2013oev,Chatrchyan:2013yaa}. They show a small excess of 1--2-$\sigma$ but 
remarkably consistent between the two experiments and different centre-of-mass energies (cms). Recent analyses 
including higher order effects~\cite{Baglio:2013toa,Dawson:2013lya,::2014uva} increase the discrepancy, calling for 
more efforts to better understand its origin. Apart from a plausible explanation of systematic effects due to e.g.\ 
jet veto, one can invoke BSM models where the excess can be explained by the production of supersymmetric (SUSY) particles: 
charginos~\cite{Curtin:2012nn}, stops~\cite{Rolbiecki:2013fia}, or sleptons~\cite{Curtin:2013gta}.  

We would also like to note that the SM measurements of $W^{\pm}$~\cite{Aad:2011dm,Chatrchyan:2014mua,Chatrchyan:2013tia}, 
$Z$~\cite{Aad:2011dm,Aad:2013iua,Chatrchyan:2014mua,Chatrchyan:2013tia}, 
$W^{\pm}Z$~\cite{Aad:2012twa,ATLAS:2013fma,CMS:2013qea} and $ZZ$~\cite{Aad:2012awa,ATLAS:2013gma,CMS:2013hea} 
production show very good agreement with the theoretical
predictions. This hints that a solution cannot be a simple re-normalisation of the cross-section 
but requires a non-trivial explanation.

More recently, a pair of \ATLAS\, SUSY searches have shown intriguing excesses in soft lepton and missing energy channels
that may also support the BSM interpretation. The first, a SUSY search for squarks and 
gluinos~\cite{ATLAS-CONF-2013-062} has a $\sim2.5$-$\sigma$ excess, in a signal region that looks for a 
pair of particularly soft muons and missing energy. Secondly, \ATLAS\, has released a tri-lepton search~\cite{Aad:2014nua} for electroweak states
where a number of different signal regions display small excesses: 1.6, 1.9 and 2.1-$\sigma$.

In this paper we focus on a simplified model with a light stop, chargino and two neutralinos and ask whether
the above excesses can be explained in a consistent way without violating any of the vast number of SUSY searches 
that have been performed. In particular we concentrate on a model with a stop that is heaviest, a wino-like neutralino 
and associated charginos that are slightly lighter and a bino-like lightest supersymmetric particle (LSP). We thus 
find that with stop production the resulting decay chain dominates,
\begin{equation}\label{eq:decay_stop}
 \sstop \to \cha{1}\; b \to \neu{1}\; W^{\pm(*)}\; b \to \neu{1}\; \ell^{\pm} \; \nu \; b\;,
\end{equation}
where the $^{(*)}$ refers to the fact that the $W^{\pm}$ may only be present as an off-shell state if the mass 
splitting, $m_{\cha{1}} - m_{\neu{1}} < m_{W^{\pm}}$. 

The stop pair production, $pp \to \sstop\sstop^*$, can therefore significantly contribute to the final state 
used for $W^+W^-$ cross section measurements, two opposite sign leptons and missing transverse energy. We choose  
small mass difference between the stop and chargino, $\mstop{1} - \mcha{1} = 7\gev$, to help to avoid the jet veto used in $W^+W^-$ 
studies. In addition, the small mass splitting only produces soft $b$-jets and thus makes the model 
less strongly constrained by direct stop searches. The final state also contributes in e.g.\ the signal regions of SUSY 
searches in the di-lepton channel~\cite{Aad:2014vma,ATLAS-CONF-2013-062} or control regions of Higgs boson 
measurements~\cite{ATLAS-CONF-2013-030,ATLAS-CONF-2013-031}. 

Along with the stop and chargino, the model also contains two neutralinos that may have significant cross sections. Therefore the 
additional production processes, $pp \to \chapm{1}\neu{2}$ and  $pp \to \chap{1}\cham{1}$, should also be 
taken into account. In particular, contributions form the first process may become visible in the SUSY tri-lepton search~\cite{Aad:2014nua}
mentioned above with the following decays,
\begin{eqnarray}
  \cha{1}\; & \to & \neu{1}\; W^{\pm(*)}\; \to \neu{1}\; \ell^{\pm} \; \nu \;, \label{eq:decay_cha} \\
  \neu{2}\; & \to & \neu{1}\; Z^{(*)}\; \to \neu{1}\; \ell^+ \; \ell^- \;.\label{eq:decay_chi}
\end{eqnarray}
Therefore, the present study significantly enhances the scope of the 
previous analysis~\cite{Rolbiecki:2013fia} that focused purely on the $W^+W^-$ cross-section measurement.

Of course it is important to not only include measurements that display excesses and bias the fit for new physics. Therefore
we include all current \ATLAS\, searches that are sensitive to the model in question. Particularly important are the searches for 
stop and bottom squarks \cite{Aad:2014qaa,ATLAS:2013pla}, squarks and gluinos with soft 
leptons \cite{ATLAS-CONF-2013-062} and electroweak searches \cite{Aad:2014vma,Aad:2014nua}. Every signal region in these 
searches is included in a global fit to provide a fair comparison with the SM. In addition, the regions in 
our model space that are excluded by these searches are given in the final results.

One should also note that additional constraints may be present in a model with light top 
quark partners. For example, such models may contain extra charge- or colour-breaking
vacua \cite{Camargo-Molina:2014pwa} but since we only work with a simplified model this
is beyond the scope of the current work.

The paper is organized as follows. In section~\ref{sec:2} we introduce tools used in the analysis and briefly discuss included \LHC\, searches. In section~\ref{sec:3} 
we provide details of the simplified model and methodology of the parameter scan. The results are described in section~\ref{sec:4} followed by a discussion in section~\ref{sec:5}. Finally we summarize our findings in section~\ref{sec:conclusions}.

\section{Tools and measurements\label{sec:2}}

The idea of the study is to provide a global fit to the SUSY model in question and compare
this to the SM. In order to perform this task we make use of studies implemented in the experimental 
re-casting tools \ATOM~\cite{ATOM} and \CheckMATE~\cite{Drees:2013wra}. Both tools make use of a fast detector implementation, 
\Rivet~\cite{Buckley:2010ar} for \ATOM \footnote{\ATOM\, includes experimental efficiencies and smearing modifications
to \Rivet\, that allow it to perform like a fast detector simulation.} and \Delphes \footnote{\CheckMATE\, includes
a detailed re-tuning of \Delphes\, to match the latest \ATLAS\, experimental setup.} \cite{deFavereau:2013fsa} 
for \CheckMATE. Both setups use \Fastjet\,\cite{Cacciari:2011ma} with the anti-$k_T$ 
algorithm~\cite{Cacciari:2005hq,Cacciari:2008gp} for all analyses. Additional tools were used to calculate 
the kinematical variable $M_{T2}$~\cite{Lester:1999tx,Barr:2003rg,Cheng:2008hk}.

All SUSY events are generated with \Herwig~\cite{Bahr:2008pv,Bellm:2013lba}. For stop production, these events are normalised by the 
NLO+NLL cross section using \NLLFast~\cite{Beenakker:1996ch,Beenakker:1997ut,Beenakker:2009ha,Beenakker:2010nq}, while 
for chargino and neutralino production, we use \Prospino~\cite{Beenakker:1999xh}. For all production processes we use the CTEQ NLO
parton distribution functions \cite{Nadolsky:2008zw}.

The \Herwig\, event generator used in this study does not contain initial state radiation (ISR) calculated at the matrix element 
level but instead relies on the parton shower as an approximation to this radiation. One may worry that if the signal regions
require hard ISR jets, we may underestimate the number of events in these regions. However, none of the studies considered 
actually requires a particularly hard ISR jet for the relevant signal regions. The `soft di-muon' channel in 
the ATLAS squark and gluino search \cite{ATLAS-CONF-2013-062} requires a jet with $p_T>70$~GeV but this is still an momentum range 
where we expect the parton shower to accurately model radiation since it is well below the typical production centre of mass energies 
for our model (at least 400~GeV).

The accuracy of the parton shower has been previously studied \cite{Dreiner:2012sh, Dreiner:2012gx} and it was found
that the \Herwig\, parton shower was within the prediction for matrix element radiation as long as the hardest jet $p_T <300$~GeV.
Only once we go into the tails of very hard jet radiation does \Herwig\, fail badly. However, the event rate in these 
tails is too small to contribute to any of the signal regions anyway. 

As stated in the introduction we include the current $W^+W^-$ measurements from both \ATLAS\, and \CMS\,
along with all \ATLAS\, searches that may be sensitive. The rational behind only choosing \ATLAS\, 
searches is that both \ATOM\, and \Checkmate\, have been tuned to \ATLAS\, but the \CMS\, tunes have not
yet been completed. In addition, it appears that the relevant \CMS\,
searches\footnote{While our study was in the last preparation stages a tri-lepton \CMS\, analysis \cite{Khachatryan:2014qwa} was made
public. The study also shows an excess in the data that may be compatible with our model and we hope to include
this soon.} \cite{Chatrchyan:2013xna,Chatrchyan:2012paa,CMS:2013dea} are less sensitive to the model we propose here. All the analyses
are listed in table~\ref{tab:exp_analyses} and below we introduce the different analysis classes in more detail.

Throughout the study, all SM backgrounds are taken from the respective experimental 
publication. However, to validate the various analyses we have generated the dominant 
SM backgrounds and our results match the experimental results very closely and always to better than 10\%.

\begin{table}\begin{center}
\renewcommand{\arraystretch}{1.}
\begin{tabular*}{1.0\textwidth}{@{\extracolsep{\fill} }lrrrrrr} \toprule
 
 Description          						& $\sqrt{s}$  		& Luminosity    & Number 		& Refs. 	\\ 
								& [TeV]			&  [$\ifb$] 	&   of SR		&    		   \\ \midrule
 \ATLAS\, $W^+W^-$   						& $7$		 	& $4.6$   	& $1$ 			& \cite{ATLAS:2012mec}             \\ 
 \CMS\, $W^+W^-$  						& 7 			& $4.9$   	& $1$ 			&  \cite{Chatrchyan:2013yaa}             \\ 
 \CMS\, $W^+W^-$  						& 8	  		& $3.5$   	& $1$			&  \cite{Chatrchyan:2013oev}      \\  
 \ATLAS\, Higgs    				 		& 8			& $20.7$	& $2$		   	&  \cite{ATLAS-CONF-2013-031}    \\ 
 \ATLAS\, Electroweak (2 $\ell$)     				& 8		  	& $20.3$   	& $13$		    	&   \cite{Aad:2014vma} 	 \\
 \ATLAS\, $\tilde{q}$ and $\tilde{g}$ (1-2 $\ell$)  		& 8		 	& $20.1$	& $19$			&   \cite{ATLAS-CONF-2013-062}   \\
  \ATLAS\, $\tilde{q}$ and $\tilde{g}$ razor (2 $\ell$)  	& 8		 	& $20.3$	& $6$			&   \cite{ATLAS-CONF-2013-089}   \\
 \ATLAS\, Electroweak (3 $\ell$) 				& 8		 	& $20.3$	& $20$			&   \cite{Aad:2014nua}        \\
 \ATLAS\, $\tilde{t}$ (1 $\ell$)  				& 8		 	& $20.7$	& $8$			&   \cite{ATLAS:2013pla}   \\
 \ATLAS\, $\tilde{t}$ (2 $\ell$) 	 			& 8			& $20.3$	& $12$ 			&   \cite{Aad:2014qaa}  		 \\
  \CMS\, $W^{\pm}Z^0$ 			 			& 8			& $19.6$	& $4$ 			&   \cite{CMS:2013qea}  		 \\
   \ATLAS\, $W^{\pm}Z^0$ 	 				& 8			& $13.0$	& $4$ 			&   \cite{ATLAS:2013fma}  		 \\
 \bottomrule
 
\end{tabular*}
\end{center}
\caption{List of experimental analyses used in this study along with the centre-of-mass energy, $\sqrt{s}$, luminosity, number of 
signal regions (SR) and the corresponding references. All signal regions listed above are included in our global model fit. \label{tab:exp_analyses}}
\end{table}

\subsection{$W^+W^-$ measurements}

The first analyses that hinted at a possible excess 
were the 7~TeV $W^+W^-$ measurements from \ATLAS~\cite{ATLAS:2012mec} 
and \CMS~\cite{Chatrchyan:2013yaa} and both are included in our fit as a simple rate measurement. An early 8~TeV measurement 
with a small dataset was also produced by \CMS~\cite{Chatrchyan:2013oev} so we include this as well. Both measurements target di-lepton plus \met\, final state.

Currently, no 8~TeV measurement has yet been provided by \ATLAS\, so we are eager to find out what this result will be. However, we are able to make an early estimate
by examining an \ATLAS\, Higgs spin study~\cite{ATLAS-CONF-2013-031} where the Higgs decays to a $W^+W^-$ pair. Here $W^+W^-$ production is the
dominant background and therefore the experiment makes a control region measurement which we can use. Unfortunately, an error analysis is 
not provided for the control region and we therefore assume that the systematic error will be the same as the \CMS\, analysis (6\%). 
Considering that the \ATLAS\, result has $\sim 6$ times as much luminosity and many of the systematic errors are statistically limited
we believe that this is a conservative choice.

We furthermore include the Higgs signal region from ref.~\cite{ATLAS-CONF-2013-031} which differs from the $W^+W^-$ control 
region solely by the lepton invariant mass cut. Therefore, both regions include the full $W^+W^-$ sample 
binned in a way that could provide sensitivity to mass difference between chargino 
and the LSP. The number of expected $W^+W^-$ events in the Higgs signal region of ref.~\cite{ATLAS-CONF-2013-031} has 
been renormalized by $1.08$ with respect to the Monte Carlo (MC) prediction.\footnote{Note that 
even after this rescaling of the $W^+W^-$ background, there is still an excess in the signal region suggesting 
that the additional contribution falls more often in the low $m_{\ell\ell}$ bin. This feature would promote 
smaller mass splitting between the chargino and the LSP.} For our purpose we therefore rescale this number, assuming 
that it was caused by the contamination of $W^+W^-$ control region by the stop production. In this way we obtain a real 
size of the excess in the observed events with 
respect to the MC prediction. The precise number of events we assume along with the systematic error and 
calculated excess are shown in table~\ref{tab:chi2}.

\subsection{SUSY di-lepton search}

We include the \ATLAS\ search for chargino and slepton production in the 
di-lepton channel~\cite{Aad:2014vma} since the study could provide important limits for stop and 
gaugino production. It is also interesting to note that like in the Higgs 
spin study~\cite{ATLAS-CONF-2013-031}, the MC-predicted and observed numbers in 
the $W^+W^-$ control regions do not match well. This excess can be again neatly explained by 
the contamination coming from stop production. 

As in the Higgs study, the $W^+W^-$ background in the signal regions is found by normalising the Monte Carlo prediction 
by the excess in the control regions. In this analysis, the rescaling actually leads to a deficit in the number of 
observed events in the signal regions. We therefore follow the same procedure as for the Higgs study and remove the 
normalisation on the $W^+W^-$ background.

\subsection{Leptonic squark and gluino searches}

General SUSY searches that include signal regions looking for soft leptons may be particularly relevant for 
our model. The original motivation for these searches was to look for compressed SUSY or universal extra 
dimensions (UED) spectra.

The first analysis~\cite{ATLAS-CONF-2013-062} we consider has both single and di-leptonic signal regions. Of 
special interest to our model is that a signal region looking for a pair of soft muons displays a $\sim 2.5$-$\sigma$ 
excess (see table~\ref{tab:chi2}). However, we include all 13 signal regions into our analysis to account for the 
fact that some of these may also be relevant for our model.

The second analysis \cite{ATLAS-CONF-2013-089} is based on the so called `Razor' variable that re-parametrises the background 
into a smoothly falling distribution. SUSY is then searched for as a departure from the expected shape. No excess is seen in the 
analysis and consequently we simply use the search to constrain our model

\subsection{SUSY tri-lepton searches}

Since in our simplified model we have the mass degenerate chargino and neutralino, $\chapm{1}$ and $\neu{2}$, SUSY 
electroweak searches could also play important role. The most relevant analysis here is the tri-lepton search 
performed by \ATLAS~\cite{Aad:2014nua}. The study targets final states with three leptons and missing energy. An 
important feature of this study is the sensitivity to events with low invariant mass of an opposite-sign lepton pairs. Indeed, 
in several signal regions of low invariant mass one can see the excess of observed events that goes up to 2.1-$\sigma$, 
see table~\ref{tab:chi2}. This results in weaker than 
expected exclusion limits for chargino-neutralino associated production, notably in the mass region of $\sim 200 \gev$. Such 
an excess of events can be easily associated with a neutralino decaying to a pair of leptons 
provided the mass splitting $m_{\neu{2}}-m_{\neu{1}}$ is not too large. As we will see later this will be an 
important contribution to our analysis.

In addition, one may worry that the tri-leptons $\chapm{1}$ and $\neu{2}$ decays may enter into the 
SM $W^{\pm}Z^0$ measurement. For this reason, we also include the analysis of this final state from both \ATLAS\, \cite{ATLAS:2013fma}
and \CMS\,\cite{CMS:2013qea} into our study.

\subsection{Stop searches}

Finally, we include the ATLAS direct searches for light and medium stops with either one \cite{ATLAS:2013pla}
or two \cite{Aad:2014qaa} isolated leptons. Since the mass splitting,
$m_{\sstop}-m_{\chapm{1}}=7$~GeV, in our model is small we expect the $b$-quarks produced in this decay to be soft and 
that these searches will only provide weak constraints. We also note that in the two lepton study a control region 
measurement of $W^+W^-$ production is made that is then extrapolated to the signal region. We once again remove this 
normalisation to account for the fact that we expect the control region to be contaminated with signal.

\section{Scan\label{sec:3}}

For our scan we choose a simplified model with the following mass hierarchy:
\begin{equation}
 m_{\sstop} > m_{\cha{1}} = m_{\neu{2}} > m_{\neu{1}}\,,
\end{equation}
and set the mass difference between the stop and chargino to $m_{\sstop} - m_{\cha{1}} = 7\gev$\footnote{We
also perform a scan with $m_{\sstop} - m_{\cha{1}} = 15\gev$ and these results are briefly commented 
on in section~\ref{sec:CombAnalysis}.}. Assuming that all 
other SUSY particles are decoupled, the only allowed decays are given 
by eqs.~\eqref{eq:decay_stop}--\eqref{eq:decay_chi} with the final state branching ratios being given by the SM couplings. The 
models under investigation are then defined by the two remaining masses: $m_{\sstop}$ and $m_{\neu{1}}$. In order to find 
the set of parameters that provides the best explanation of the existing data we perform a scan in these two parameters. The 
production cross section of $\chapm{1}\neu{2}$ and $\tilde{\chi}^+_1\tilde{\chi}^-_1$ is set by assuming wino nature of $\chapm{1}$ and $\neu{2}$.  

In realistic SUSY models, several modifications of the above scenario could be possible. Firstly, instead of wino-like 
chargino and neutralino, one could have higgsino-like $\cha{1}$ accompanied by two neutralinos, $\neu{2}$ and $\neu{3}$. This 
will not change the decay patterns above and only the cross section for production of $\cha{1}\neu{2,3}$ would be reduced 
compared to the pure wino case. On the other hand, if sleptons are light --- but still heavier than the chargino --- the branching 
ratios could be significantly affected enhancing the leptonic decay modes when only three-body decays are allowed. Such a modification 
could be motivated by dark matter constraints, allowing for larger annihilation cross sections, or by the muon anomalous magnetic moment.  
Finally, right-handed sleptons lighter than the chargino will not affect its decays because of the vanishing coupling between wino and the right-handed sleptons.
In case of staus, this would require a very small left-right mixing and small higgsino component in the light chargino. The main advantage of such a modification,
if $m_{\tilde{\ell}} - \mneu{1} \sim 10 \gev$, would be dark matter relic density consistent with current measurements for the bino LSP via co-annihilation with sleptons.  

To perform the scan, we generate a grid in the $(m_{\sstop}, m_{\neu{1}})$ plane with a step size of $5\gev$ in 
the ranges $150\gev < m_{\sstop} < 300\gev$, $60\gev < m_{\neu{1}} < 300\gev$. A corresponding grid is generated for 
chargino and neutralino production. The number of generated events followed by leptonic decays varies across 
the plane according to the production cross section and expected number of signal events, in order to reduce the impact of MC 
statistical errors. Since some of the signal regions of the studies under consideration, notably ATLAS-CONF-2013-062~\cite{ATLAS-CONF-2013-062} and
ATLAS-CONF-2013-089~\cite{ATLAS-CONF-2013-089}, 
have very small acceptances, up to $2.5 \cdot 10^6$ events has been generated around the region of best fit.

The fit was performed with the following test statistics. For signal regions (or measurements) with large 
numbers of expected and observed events ($>20$)
we use the $\chi^2$ statistic,
\begin{eqnarray}
 \chi_i^2 = \frac{(n_i-\mu_i)^2}{\sigma^2_{i,{\rm stat}}+\sigma_{i,b}^2}\;, 
\end{eqnarray}
where
\begin{eqnarray}
 \mu_i = \mu_{i,b}+\mu_{i,s}\;.
 \end{eqnarray}
Here, $n_i$ is the number of observed events, $\mu_{i,b}$ is the expected number of background 
events, $\mu_{i,s}$ is the expected number of signal 
events, $\sigma_{i,{\rm stat}}$ and $\sigma_{i,b}$ are the statistical and systematic uncertainty 
on the expected number of background events for each signal region, $i$.

However, for signal regions with small numbers of expected or observed events, the $\chi^2$ statistic 
becomes unreliable and we therefore use the likelihood-ratio
test statistic~\cite{Stats},
\begin{equation}
 -2\ln \lambda_j = -2\ln \left(\frac{\int^{\infty}_0 db_j'\frac{(\mu'_j)^{n_j}e^{-\mu'_j}}{n_j!}G(\mu_j,b_j')}{\frac{(n_j)^{n_j}e^{-n_j}}{n_j!}}\right)\;.
\end{equation}
for each signal region $j$.

The systematic uncertainty on the background is included by marginalising over this error and is always assumed
to be Gaussian. Specifically, we perform an
integral where we shift $\mu_j\to\mu_j(1+f(b'))$, where $f(b')$ is drawn from a Gaussian, $G$, with standard deviation, $\sigma_{j,b}$.

The overall test statistic is then given by the summation of the small and large event signal regions,\footnote{Note that
the definition of the likelihood means that $-2\ln \lambda$ is always positive.}
\begin{equation}
 -2\ln L = \sum_{i=1}^{N_h} \chi^2_i + \sum_{j=1}^{N_l} -2\ln \lambda_j\;.
\end{equation}
Here $N_h$ refers to the number of high statistic signal regions with both $\mu_i,n_i\geq20$ whereas $N_l$ 
refers to the number of low statistic signal regions with both $\mu_j,n_j<20$.

We make the assumption that all errors are uncorrelated when we perform the fit. Seeing
as many of the detector effects, background samples and signal samples are clearly correlated
one may worry about how justified the assumption is. Unfortunately, the experiments 
often do not even provide a detailed breakdown of systematic errors in a single analysis,
let alone provide a detailed correlation matrix. Therefore we found that it is impossible 
to reliably estimate how big the correlations would be within a single analysis let alone
the different disjoint analyses we consider here. In the discussion section~\ref{sec:5},
we consider how including correlations may effect our final conclusions.

\section{Results\label{sec:4}}

Considering that current \LHC\, excesses occur in both two and three lepton signal regions, an interesting question is 
whether our stop model can simultaneously explain both sets of data. For the di-lepton channels, the dominant process 
is stop pair production followed by the decay chain, 
\begin{equation}\label{eq:decay_stop_res}
 \sstop \to \cha{1}\; b \to \neu{1}\; W^{\pm(*)}\; b \to \neu{1}\; \ell^{\pm} \; \nu \; b\;.  
\end{equation}
However, for the tri-lepton signal, the dominant channel is the electroweak production of $pp \to \chapm{1}\neu{2}$ 
with the following decays,
\begin{eqnarray}
  \cha{1}\; & \to & \neu{1}\; W^{\pm(*)}\; \to \neu{1}\; \ell^{\pm} \; \nu \;, \label{eq:decay_cha_res} \\
  \neu{2}\; & \to & \neu{1}\; Z^{(*)}\; \to \neu{1}\; \ell^+ \; \ell^- \;.\label{eq:decay_chi_res}
\end{eqnarray}
Thus, we explore the fits to the two and three leptons channels separately before providing a combination.

\subsection{Di-lepton channels}

\begin{figure}
 \centering \vspace{-0.0cm}
 \includegraphics[width=0.9\textwidth]{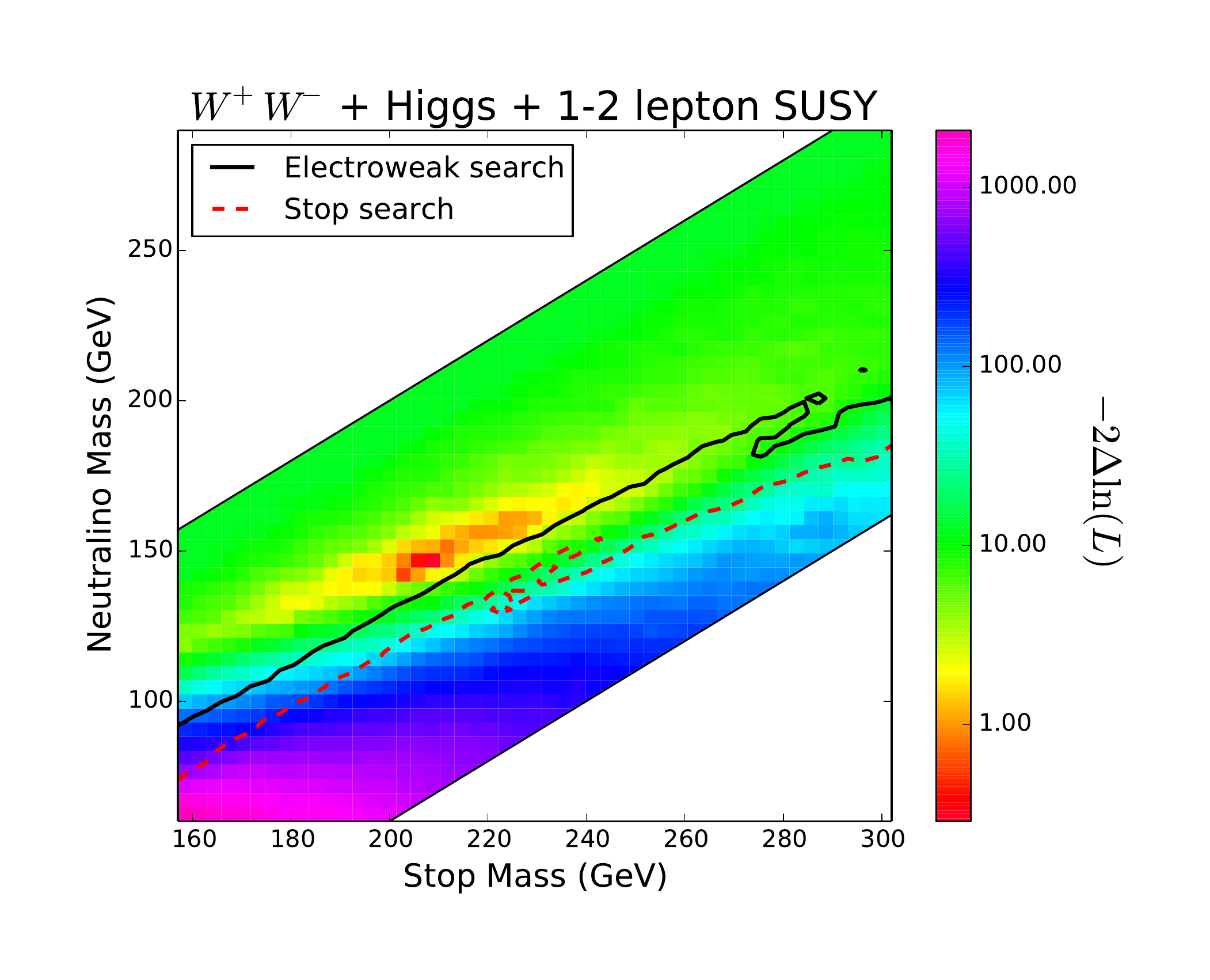}
 \caption{The distribution of $-2\ln L$ as a function of the masses of the stop, $m_{\tilde{t}_1}$, 
 and lightest neutralino, $m_{\tilde{\chi}^0_1}$. All of the signal regions given in the
 $W^+W^-$ measurements, any searches with di-lepton final states and the stop searches, table~\ref{tab:exp_analyses} are 
 included in the fit. Also 
 shown are the 95\% CLs
 exclusion lines given by the dedicated \ATLAS\, search for SUSY electroweak states \cite{Aad:2014vma} and the 
 dedicated \ATLAS\, stop search \cite{Aad:2014qaa}.
 \label{fig:StopNeut_7GeV}}
\end{figure}

For the di-lepton fit\footnote{We also include the single lepton 
stop analysis \cite{ATLAS:2013pla} and single lepton squark and gluino 
analysis \cite{ATLAS-CONF-2013-062} here since these searches are most sensitive 
to stop production.} we include all experimental analyses listed in table~\ref{tab:chi2} with the exception
of tri-lepton search \cite{Aad:2014nua} and the $W^{\pm}Z^0$ measurements \cite{ATLAS:2013fma,CMS:2013qea}
since stop production will produce a negligible number of events with three leptons. 
  
Performing a scan as a function of $m_{\sstop}$ and $m_{\tilde{\chi}^0_1}$ while 
keeping the mass splitting $m_{\sstop} - m_{\cha{1}}=7\gev$ we find that the stop model fits the 
current data significantly better than the SM as can be seen in 
figure~\ref{fig:StopNeut_7GeV}. The best fit point is found to be,
\begin{eqnarray}
    m_{\sstop} = 207^{+40}_{-30} \;\mathrm {GeV},   \\
    m_{\tilde{\chi}^0_1} = 145^{+25}_{-10} \;\mathrm {GeV}.
\end{eqnarray}
Compared to the SM we find a reduction in the log-likelihood, $-2\ln L = 11.9$, which corresponds to $\sim$ 3.0-$\sigma$
once the extra degrees of freedom in the fit are considered.

\begin{figure}
 \centering \vspace{-0.0cm}
 \includegraphics[width=0.9\textwidth]{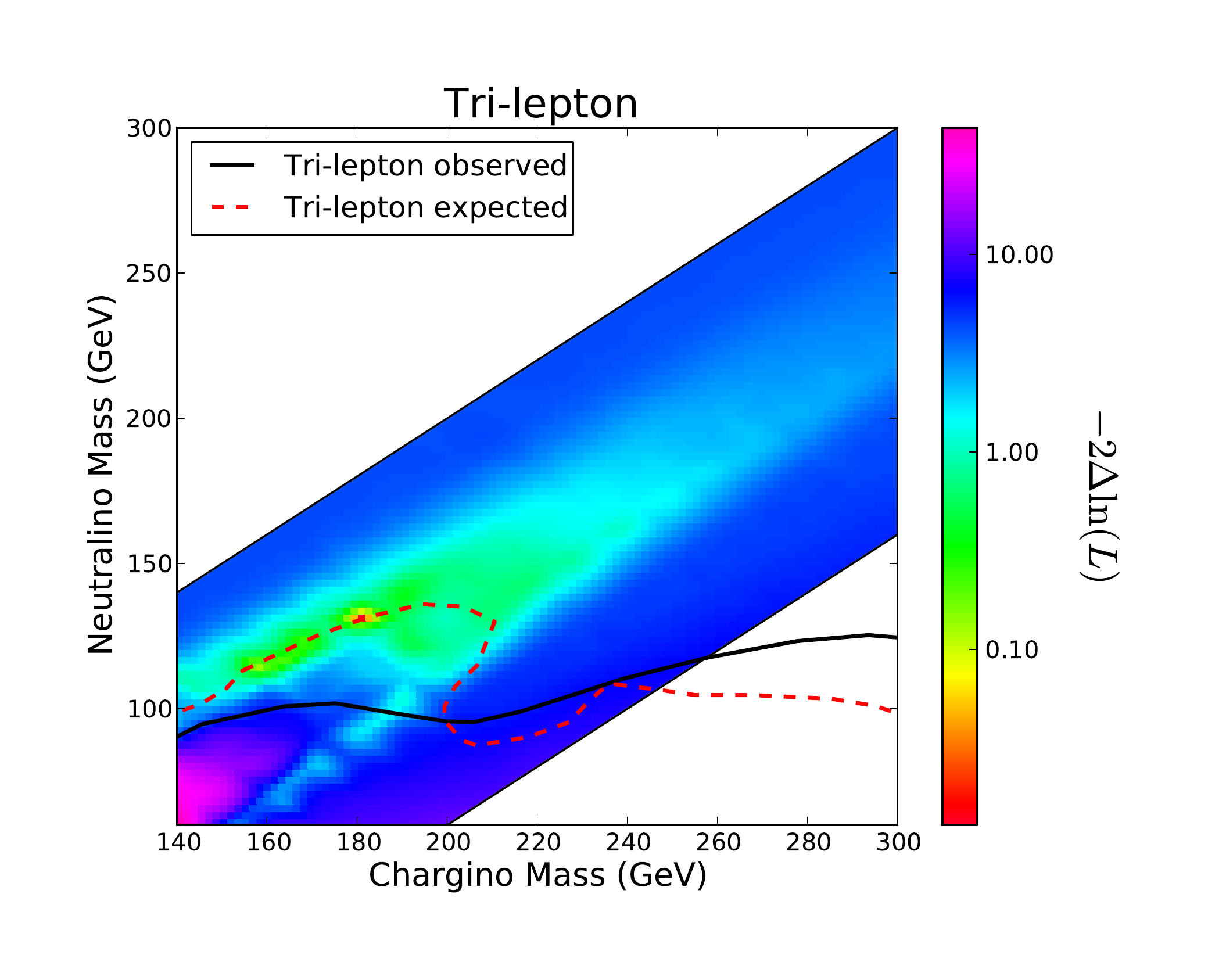}
 \caption{The distribution of $-2\ln L$ as a function of the masses of the lightest chargino, $m_{\tilde{\chi}_1^{\pm}}$, 
 and lightest neutralino, $m_{\tilde{\chi}^0_1}$. {All signal regions in the} \ATLAS\, tri-lepton search for charginos and neutralinos
 are used in the fit \cite{Aad:2014nua}. Also shown are the observed and expected 
 95\% CLs exclusion lines for the same search.
 \label{fig:CharNeut}}
\end{figure}

In addition we plot the exclusion limits from the most constraining \ATLAS\, supersymmetric searches in figure~\ref{fig:StopNeut_7GeV}. We 
find that the electroweak search for two leptons \cite{Aad:2014vma} best constrains our model but cannot exclude the best fit 
point. However, since the exclusion probes mass splittings between the $m_{\sstop}$ and $m_{\tilde{\chi}^0_1}$ only $\sim 15$~GeV
larger than our preferred region, we believe that there is a good chance that these searches can probe the model in the near future.
  
  \begin{figure}
 \centering \vspace{-0.0cm}
 \includegraphics[width=0.9\textwidth]{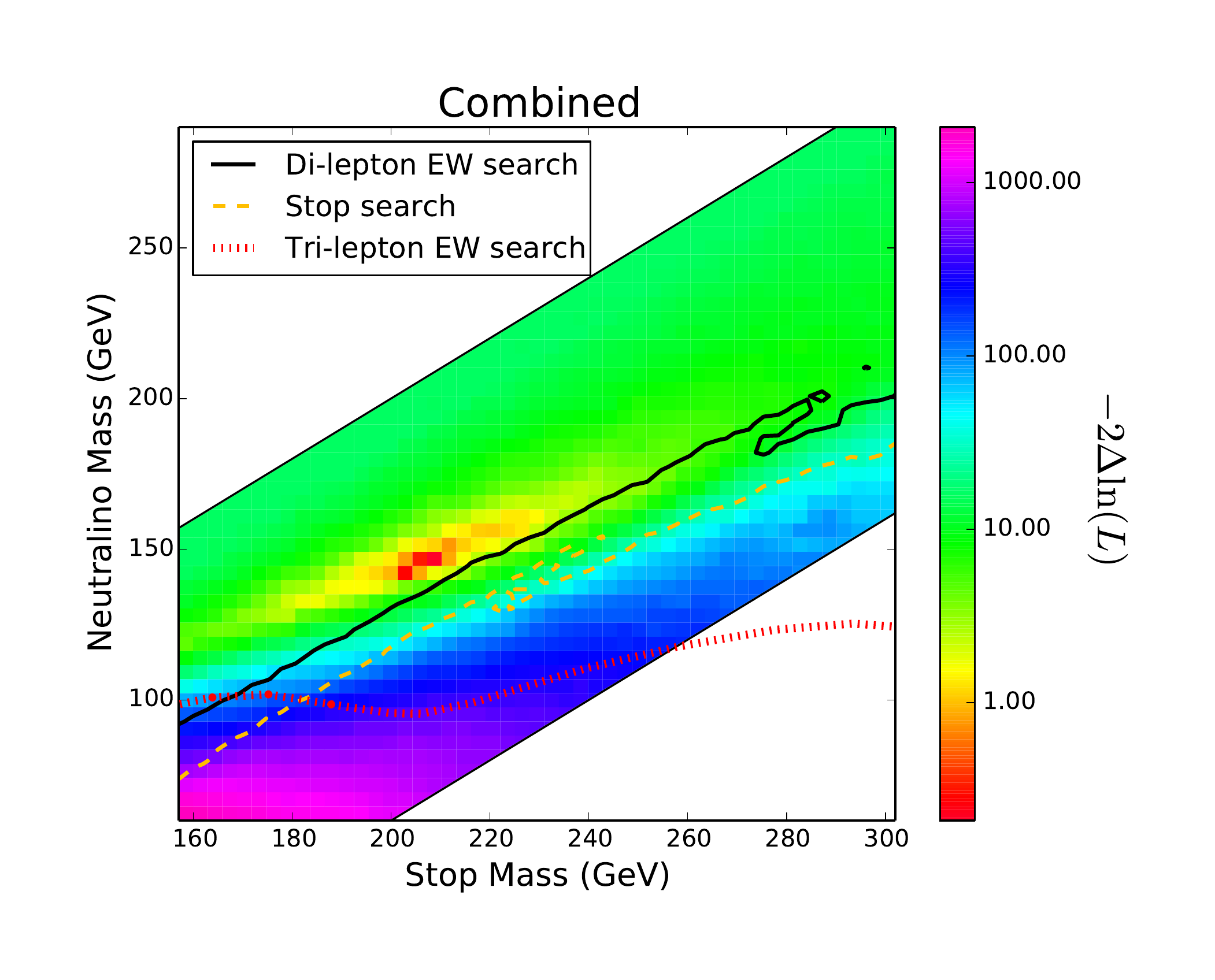}
 \caption{The distribution of $-2\ln L$ as a function of the masses of the stop, $m_{\tilde{t}_1}$, 
 and lightest neutralino, $m_{\tilde{\chi}^0_1}$. All of the signal regions given in the
 $W^+W^-$ measurements, di-lepton and tri-lepton searches, table~\ref{tab:exp_analyses} are included in the fit. Also shown are 
 the 95\% CLs exclusion lines given by the dedicated \ATLAS\, di-lepton \cite{Aad:2014vma}, 
 tri-lepton \cite{Aad:2014nua} and stop \cite{Aad:2014qaa} searches.
 \label{fig:Combined_WZ_7GeV}}
\end{figure}
  
\subsection{Tri-lepton channels}

For the tri-lepton fit include the dedicated \ATLAS\, search \cite{Aad:2014nua} and the $W^{\pm}Z^0$ measurements to 
concentrate on the electroweak production
of $\chapm{1}\neu{2}$ followed by the decay chains given in eq.~\eqref{eq:decay_cha_res} and \eqref{eq:decay_chi_res}. We scan over the 
$\chapm{1} = \neu{2}$ and $\neu{1}$ mass space and remind readers that we consider $\chapm{1},\neu{2}$ to be purely 
wino and $\neu{1}$ bino.

The best fit point for our model is to be found with the following parameters,
\begin{eqnarray}
   m_{\chapm{1}} = m_{\neu{2}} = 180^{+90}_{-50} \;\mathrm {GeV},   \\
   m_{\neu{1}} = 130^{+60}_{-30} \;\mathrm {GeV}.
\end{eqnarray}
Compared to the SM, including $\chapm{1},\neu{2}$ production reduces the log-likelihood
of the fit by $4.3$ which corresponds to $1.6$-$\sigma$ once the extra degrees of freedom in the fit are considered.

As can be seen in figure~\ref{fig:CharNeut}, the best fit point lies well above the current 
experimental exclusion. However, one should notice that the expected exclusion goes almost 
exactly through our best fit point. Consequently, the current experimental analysis is sensitive to
the model we propose. Indeed, if the excess present in the data is only due to a statistical fluctuation,
one would expect the point to be ruled out in the forthcoming \LHC\, run at 14~TeV. The breakdown
of signal regions with significant contributions from our model point is given in table \ref{tab:ATLAS_3l} 
of appendix \ref{app:SRbreakdown}.

Finally we also note that at the best fit point our model has a mass 
difference $m_{\chapm{1},\neu{2}} - m_{\neu{1}} \sim 50$~GeV. Consequently, our model
contributes negligibly to $W^{\pm}Z^0$ measurements since the invariant mass of the leptons produced 
in the $\neu{2}$ decay lies outside the normal mass window defined for the $Z^0$.

\subsection{Combined analysis}
\label{sec:CombAnalysis}

\begin{table}\begin{center}
\renewcommand{\arraystretch}{1.}
\begin{tabular*}{1.0\textwidth}{@{\extracolsep{\fill} }llrrrrr} \toprule
 
 Study               					& SR       	& Obs	 	& Exp      		&  SM  	& Best 		& Best \\ 
							& 		&    		&    			&   s.d. 		&fit exp	& fit s.d      \\ \midrule\midrule
 \ATLAS\, $W^+W^-$ (7~TeV) \cite{ATLAS:2012mec}  	& Combined 	& $1325$   	& $1219 \pm 87$ 	& 1.1-$\sigma$   &     $119$  &   0.1-$\sigma$           \\ \midrule
 \CMS\, $W^+W^-$ (7~TeV) \cite{Chatrchyan:2013yaa} 	& Combined 	& $1134$   	& $1076 \pm 62$ 	& 0.8-$\sigma$   &     $89$  &   0.4-$\sigma$            \\ \midrule
 \CMS\, $W^+W^-$ (8~TeV)  \cite{Chatrchyan:2013oev}  	 & Combined 	& $1111$   	& $986 \pm 60$ 		& 1.8-$\sigma$   &     $83$  &  0.6-$\sigma$            \\ \midrule
 \ATLAS\, Higgs 					& $WW$ CR  	& $3297$   	& $3110 \pm 186$	& 0.9-$\sigma$   &    	$374$	&   0.9-$\sigma$\\  
 \cite{ATLAS-CONF-2013-031} 				& Higgs SR	& $3615$	& $3288 \pm 220$	& 1.4-$\sigma$   &	$501$	&  0.6-$\sigma$         \\ \midrule
 \ATLAS\, $\tilde{q}$ and $\tilde{g}$    		& Di-muon	& $7$		& $1.7 \pm 1$   	& 2.5-$\sigma$   &   $2.7$ &    1.2-$\sigma$   \\
 (1-2 $\ell$)  \cite{ATLAS-CONF-2013-062} 			&   		&    		&    			&    		&    & \\\midrule
 \ATLAS\, Electroweak               			& SR0$\tau$a01  & $36$   	& $23 \pm 4$    	& 2.1-$\sigma$	 &   $2.8$ &   1.6-$\sigma$ \\
 (3 $\ell$) \cite{Aad:2014nua}				& SR0$\tau$a06  & $13$		& $6.6 \pm 1.9$		& 1.9-$\sigma$   &   $1.5$ & 1.4-$\sigma$  \\
 \bottomrule
 
\end{tabular*}
\end{center}
\caption{Significant excesses present in the dataset under investigation. We give the signal region (SR)
of interest, the observed number of events (Obs), the expected number of events (Exp) along with 
the associated systematic error. The SM standard deviation (s.d.) for each signal region is also
given along with the expected number of events for our model best fit and the associated 
standard deviation. We only show the systematic 
error on the signal regions but the statistical errors are also included in the fit. \label{tab:chi2}}
\end{table}

One should already notice that the best-fit points for both $\sstop\sstop^*$ production and $\chapm{1}\neu{2}$ production
lie very close and well within the 1-$\sigma$ regions. We therefore perform a combined fit again as a 
function of $m_{\sstop}$ and $m_{\tilde{\chi}^0_1}$ whilst keeping the mass splitting $m_{\sstop} - m_{\chapm{1},\neu{2}} = 7$~GeV.

The result is shown in figure~\ref{fig:Combined_WZ_7GeV} and we find the best fit point for our model to be,
\begin{eqnarray}
    m_{\sstop} = 202^{+35}_{-25} \;\mathrm {GeV},   \\
    m_{\tilde{\chi}^0_1} = 140^{+25}_{-15} \;\mathrm {GeV}.
\end{eqnarray}
Comparing with the SM we find a reduction in the log-likelihood of 15.4
which corresponds to 3.5-$\sigma$ once the extra degrees of freedom in the fit are considered. In table~\ref{tab:chi2}
we show the breakdown of the different signal regions that display significant excesses and the improvement in the 
standard deviation due to our model. We see that all the $W^+W^-$ measurements (including the Higgs) present a 
compelling improvement in the agreement with data.

The di-lepton measurements dominate the fit and thus the improvement in the tri-lepton signal regions is less stark.
However, we still improve the compatibility with data including a reduction in the standard deviation from 2.1-$\sigma$ 
$\to$ 1.6-$\sigma$ for the signal region `SR0$\tau$a01'.

Another question one may ask is how does the fit quality change as the mass splitting between the stop and 
chargino is increased? If we increase the mass splitting to 15~GeV (compared to the 7~GeV presented), we find that the best 
fit point still has $m_{\tilde{t_1}}=202$~GeV. However, the minimum of the log-likelihood increases by $\sim 1$, since the 
direct stop searches now become more sensitive to the model. This is due to the $b$-quark in the $\tilde{t}_1 \to b \chapm{1}$
decay being harder and thus easier to reconstruct. As the mass difference is increased more, we expect the fit quality to
deteriorate further.

\section{Discussion\label{sec:5}}

\begin{figure}
 \centering \vspace{-0.0cm}
 \subfigure[]{\includegraphics[width=0.57\textwidth]{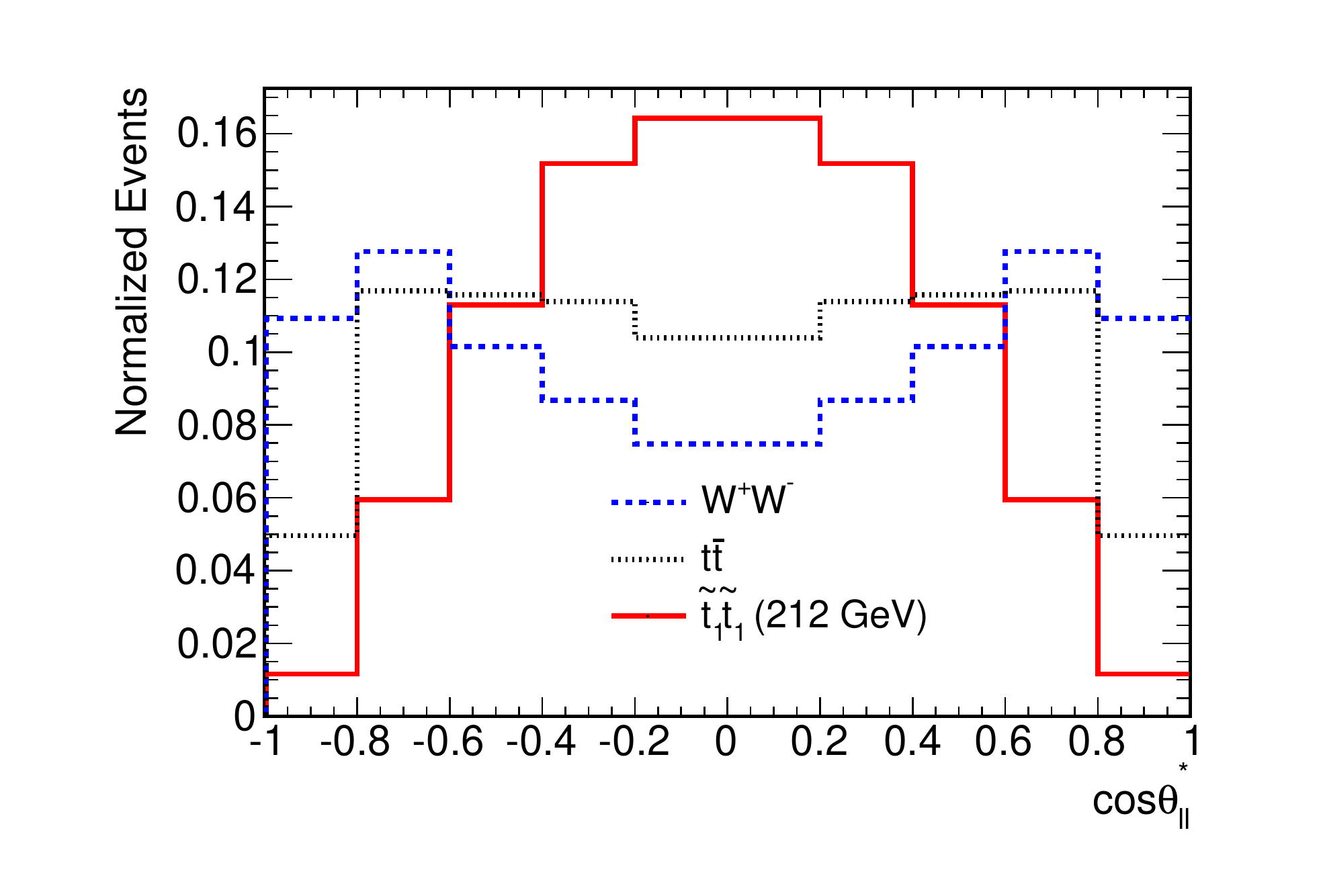}}
 \subfigure[]{\includegraphics[width=0.42\textwidth]{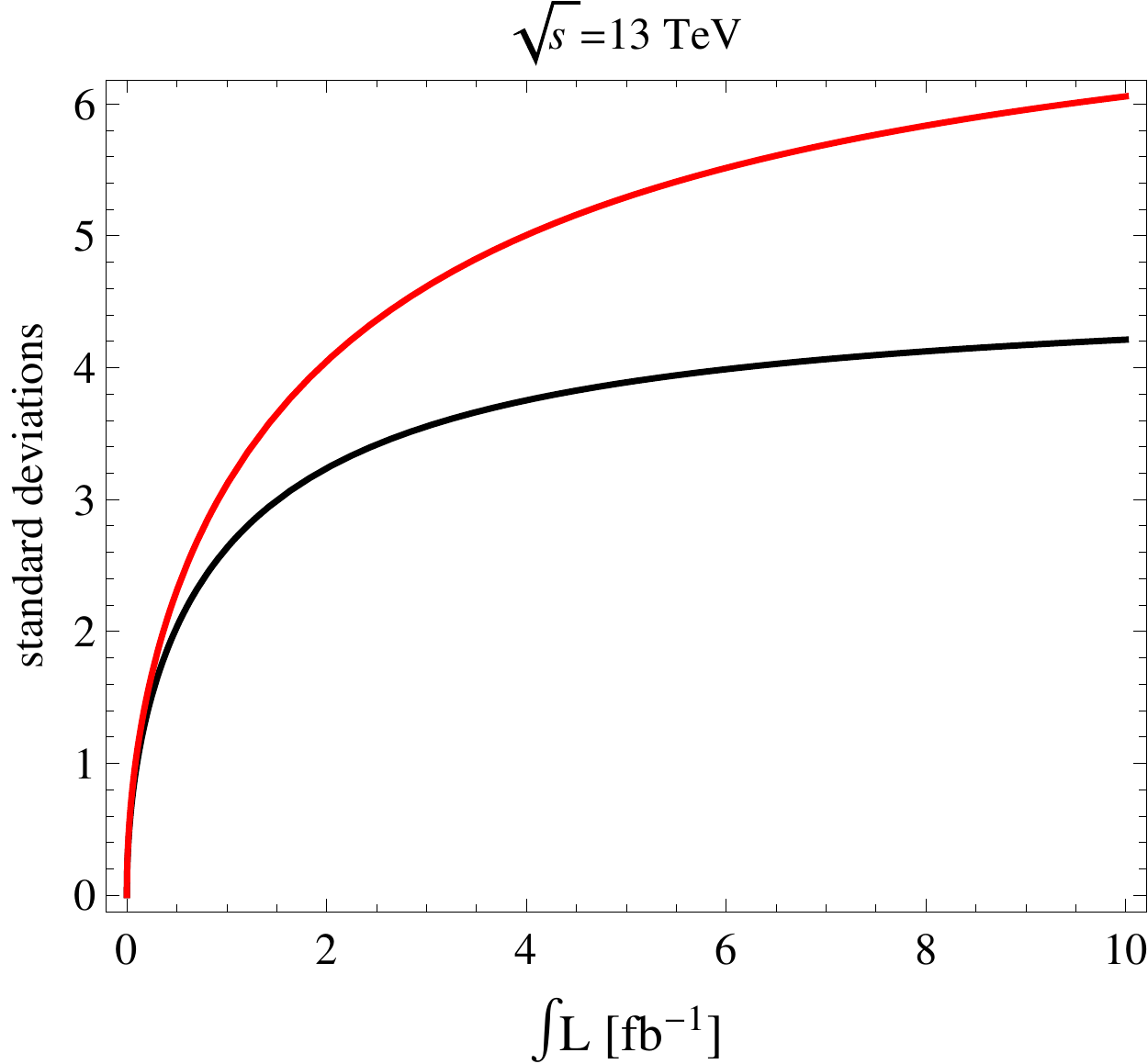}
	      \put(-60,98){\scriptsize$\sqrt{\hat{s}}_\text{min}>0$}
	      \put(-75,130){\scriptsize$\sqrt{\hat{s}}_\text{min}>150$~GeV}}
 \caption{\textbf{(a)} The distribution of $\cos\theta_{\ell\ell}^*$ for leptons produced from 
 SM $W^+W^-$ and $t\bar{t}$ events, 
 and from $\sstop\sstop^*$ events for our best fit point. \textbf{(b)} The significance of 
 distinguishing the SM-only and SM+$\sstop\sstop^*$ case as a function of an 
 integrated luminosity using the asymmetry eq.~\eqref{eq:asymmetry}. The red curve shows the 
 significance with the cut eq.~\eqref{eq:sqrtshat}, while the black curve without the $\sqrt{\hat{s}}_\text{min}$ cut. 
 \label{fig:costh}}
\end{figure}

If the observed excess in $W^+W^-$ measurements is confirmed with a larger significance in the 
full 8~TeV dataset, it will be important to confirm its BSM origin. As argued in 
ref.~\cite{Rolbiecki:2013fia} a useful observable~\cite{Barr:2005dz,Diener:2009ee,MoortgatPick:2011ix} for that purpose is  
\begin{equation}\label{eq:costhstar}
\cos\theta_{\ell\ell}^* = \tanh\left(\frac{\Delta \eta_{\ell\ell}}{2}\right) \;, \qquad \Delta\eta_{\ell\ell}= \eta_{\ell_1} - \eta_{\ell_2}\;,
\end{equation}
where $\Delta  \eta_{\ell\ell}$ is a difference of the pseudo-rapidities between the 
leading and the trailing lepton. This variable is a cosine of the polar angle of the leptons with 
respect to the beam axis in the frame where the pseudo-rapidities of the leptons are equal 
and opposite, as discussed in ref.~\cite{Barr:2005dz}. Being a function of the difference 
of pseudo-rapidities, it is longitudinally boost-invariant. Following a discussion in 
ref.~\cite{Rolbiecki:2013fia} we extend its sensitivity by requiring 
\begin{equation}\label{eq:sqrtshat}
\sqrt{\hat{s}}_\text{min} = \sqrt{E^2-P_z^2} + \mathmet > 150 \gev,
\end{equation}
with $E, P_z$ being the total energy and longitudinal momentum of the reconstructed leptons. The resulting distribution 
is shown in figure~\ref{fig:costh}(a). In figure~\ref{fig:costh}(b) we show the evolution of the significance
as a function of luminosity collected at 13~TeV for the asymmetry defined as
\begin{equation}\label{eq:asymmetry}
\mathcal{A} = \frac{ N(|\cos\theta_{\ell\ell}^*|> 0.5) - N(|\cos\theta_{\ell\ell}^*| < 0.5)}{N_{\mathrm{tot}}}\;,
\end{equation}
where $N(\ldots)$ is the number of events fulfilling the respective condition. The value of the asymmetry for stop signal alone is $\mathcal{A}_{\sstop} = -0.52$, for the SM contribution $\mathcal{A}_{\mathrm{SM}} = 0.12$ and for the sum of stop and SM contributions $\mathcal{A}_{\mathrm{SM}+\sstop} = -0.04$.
After $\sim $5~\invfb\ a 5-$\sigma$ discrimination between SUSY and the
SM should be achievable while with the full 8~TeV data sample the expected significance is 3-$\sigma$. In addition a shape analysis of the $\cos\theta_{\ell\ell}^*$ distribution can further enhance 
significance of the discovery~\cite{Rolbiecki:2013fia}.

Soft $b$-jets coming from stop decays could also be exploited to distinguish the stop contribution 
from genuine $W^+W^-$ production. A detailed analysis of $t\bar{t}$ background would be required 
in this case. This technique would require a detailed detector simulation and is beyond the scope 
of this study. We note however, that observation of such a signal would clearly point to the BSM 
partners of the third generation quarks. Another possible method is to study correlations
of tagging jets \cite{Buckley:2014fqa} that can 
be different between stop production
and background processes.

With a higher integrated luminosity and production cross section when the \LHC\, moves
to 13~TeV one could envisage a study of the di-lepton distribution from the neutralino $\neu{2}$ decays. 
This would enable a direct measurement of the mass difference, $\mneu{2} - \mneu{1}$. Together 
with the cross section measurement a more precise measurement of absolute masses could be expected.

A final enticing prospect is the hope that other SUSY states could be in reach of the \LHC. The naturalness
arguments that predict light stops also dictate that the gluino should not be much heavier than the current
bounds. Perhaps with the extended centre of mass energy, these states will become fully accessible.
Related to the question of naturalness is the mass of the higgsinos. In our model we chose the bino to be lightest
SUSY particle and the wino to be close in mass to the stop. However, we believe either of these states could be 
replaced by higgsinos and the fit would proceed with a similar result. Another possibility is that the higgsinos
could lie just above the stop mass and in this case they would have no effect on the phenomenology presented here.

In this study we chose what we think is one of the simplest SUSY model in order to explain the 
current \LHC\, excesses by only
varying the stop and LSP mass. In doing so, we took 
the `canonical' choice of a bino LSP and wino NLSP and a primarily right-handed stop to avoid 
any possible constraints from Higgs data or
electroweak precision tests~\cite{Drees:1990dx,Heinemeyer:2013dia}.
We anticipate that this simple 
choice may attract criticism since it will not predict 
the correct dark matter relic density. However, 
we believe that there are two trivial solutions to this problem that will only minimally alter the \LHC\, phenomenology.
Firstly, by allowing a generic mixing of the parameters of the neutralino and chargino mass matrix, 
a `well tempered' LSP~\cite{ArkaniHamed:2006mb} can be achieved.
Secondly, introducing an extra particle that lies close in mass to the LSP (for example a NMSSM 
singlino~\cite{Barger:2005hb} or a right-handed slepton) can lead to effective co-annihilation~\cite{Kraml:2008zr,Kim:2014noa}.

In addition, one can also question whether the outstanding problem of the anomalous magnetic moment 
of the muon could be addressed. Essentially, we would 
require light sleptons in the vicinity of the current best-fit point. Simply by placing their mass 
above the chargino they would clearly have a non-negligible impact on the branching ratios of chargino. 
One of the consequences could be shifting the preferred chargino mass higher since with the enhanced 
leptonic branching ratio the lower cross section would suffice reproduce the current excess. If lighter, 
they could also play a role in enhancing annihilation cross section of the LSP and also decays of charginos 
via an intermediate slepton.

The reason we chose not to include these effects is that in the general MSSM, the parameter space allows a 
huge range of possibilities that the current 
data cannot distinguish. We worried that adding (at least two) additional parameters to fit two 
measurements may detract and distract from the fact
that the current \LHC\, data can be fit with a far simpler model.

One should also mention of course that on a phenomenological level, there is no reason to conclude that a 
SUSY model fits the current data best 
and it is likely that many other models will also be able match satisfactorily.

\section{Conclusions\label{sec:conclusions}}
In this study we have analysed a simplified model with light stop quarks, charginos and neutralinos. A 
detailed scan of the available parameter space was performed by checking the compatibility of 
this model with a number of searches and measurements performed by the \LHC\, experiments. The 
most significant studies in our analysis are 
the $W^+W^-$ cross section measurements, Higgs spin measurement and SUSY tri-lepton searches. All of them show 
moderate excesses in the observed number of events.

We found that the inclusion of stop production improves the fit to the data with respect to the SM 
with a reduction in the log-likelihood of 15.5. Once the extra degrees of freedom in the fit are considered,
this corresponds to 3.5-$\sigma$ difference.  The parameter point that best fits the 
available data is given by $\mstop{1} = 202$~GeV, $\mcha{1} = \mneu{2} = 195$~GeV, 
and $\mneu{1} = 140$~GeV. The production of light stops, $pp \to \sstop \sstop^*$, contributes to di-lepton observables, 
while chargino-neutralino production, $pp\to \chapm{1}\neu{2}$ helps explain the excess in the SUSY tri-lepton search.

An additional observable based on the rapidities of the final state leptons could further increase the significance of the discovery. It provides a strong hint about the BSM nature of the excess in $W^+W^-$ cross section measurements.

While the available measurements are not conclusive as whether new physics has indeed been observed at the current \LHC\, run, 
the several experimental hints presented here point to an exciting possibility that requires 
further investigation. It is worth noting that several different studies point to the same region in parameter 
space that is also not excluded by the other available searches. 

The situation may become more clear when the final measurement of $W^+W^-$ production cross section 
at $\sqrt{s} = 8 \tev$ is released by \ATLAS\, and \CMS. However, the final answer will be provided by 
the future measurements with the \LHC\, at 13~TeV. With a higher centre-of-mass energy and an integrated luminosity 
increased by an order of magnitude, the nature of the excesses, if they persist, can be studied in detail. 

\section*{Note}

Simultaneous to our work, ref.~\cite{Curtin:2014zua} have also investigated the $W^+W^-$ excess with SUSY models and results
similar to the ones presented here.

\acknowledgments
We would like to thank the crucial help of Daniel Schmeier in developing \CheckMATE\, and helping with many of the analyses 
used here. We would also like to thank Matthew Dolan for pointing out errors in the way we had included 
Higgs events in the $W^+W^-$ measurements.
This work has been partially supported by the MICINN, Spain, under contract FPA2010-17747; Consolider-Ingenio  CPAN CSD2007-00042; and the European Commission under contract PITN-GA-2009-237920. JSK also thanks the
Spanish MINECO Centro de excelencia Severo Ochoa Program under grant
SEV-2012-0249. The work of K.S. was supported in part by
the London Centre for Terauniverse Studies (LCTS), using funding from
the European Research Council 
via the Advanced Investigator Grant 267352.

\appendix

\section{Signal Region Events} \label{app:SRbreakdown}

\begin{table}[h]
\begin{center}
\renewcommand{\arraystretch}{1.}
\begin{tabular*}{1.0\textwidth}{@{\extracolsep{\fill} }lccccc} 

  \multicolumn{3}{l}{\textbf{ATLAS Electroweak (3$\ell$)}} & &  \\ \midrule
Signal       	& Observed 	 	& Expected      	&  Standard Model  	& Best fit 			& Best fit\\ 
 Region 	&    events		&    events		&   standard deviation 		& events	&  standard deviation      \\ \midrule\midrule
 SR0$\tau$a01  & $36$   		& $23 \pm 4$    	& 2.1-$\sigma$	 &   $2.8$ &   1.6-$\sigma$ \\
 SR0$\tau$a02  & $5$   			& $4.2 \pm 1.5$    	& 0.8-$\sigma$	 &   $1.1$ &   0.6-$\sigma$ \\
 SR0$\tau$a03  & $9$   			& $10.6 \pm 1.8$    	& 0.7-$\sigma$	 &   $0.6$ &   0.8-$\sigma$ \\
 SR0$\tau$a04  & $9$	   		& $8.5 \pm 1.7$    	& 0.6-$\sigma$	 &   $0.7$ &   0.5-$\sigma$ \\
 SR0$\tau$a05  & $11$   		& $12.9 \pm 2.4$    	& 0.7-$\sigma$	 &   $1.4$ &   0.9-$\sigma$ \\      
 SR0$\tau$a06  & $13$			& $6.6 \pm 1.9$		& 1.9-$\sigma$   &   $1.5$ & 1.4-$\sigma$  \\
 SR0$\tau$a07  & $15$			& $14.1 \pm 2.2$	& 0.6-$\sigma$   &   $0.6$ & 0.5-$\sigma$  \\
 SR0$\tau$a08  & $1$			& $1.1 \pm 0.4$		& 0.4-$\sigma$   &   $0.2$ & 0.4-$\sigma$  \\
 SR0$\tau$a09  & $28$			& $22.4 \pm 3.6$	& 1.2-$\sigma$   &   $0.2$ & 1.1-$\sigma$  \\
 SR0$\tau$a10  & $24$			& $16.4 \pm 2.4$       & 1.6-$\sigma$	 &   $0.4$ & 1.5-$\sigma$ \\ 
 SR0$\tau$a11  & $29$			& $27 \pm 5$           & 0.8-$\sigma$	 &   $0.4$ & 0.8-$\sigma$ \\
 SR0$\tau$a12  & $8$			& $5.5 \pm 1.5$        & 1.1-$\sigma$	 &   $0.3$ & 1.0-$\sigma$ \\
 \bottomrule
 
\end{tabular*}
\end{center}
\caption{Signal regions from the ATLAS electroweak 3$\ell$ search \cite{Aad:2014nua} where the SUSY signal
contributes to the fit. We give the signal region
of interest, the observed number of events, the expected number of events along with 
the associated systematic error. The SM standard deviation for each signal region is also
given along with the expected number of events for our model best fit and the associated 
standard deviation. We only show the systematic 
error on the signal regions but the statistical errors are also included in the fit. \label{tab:ATLAS_3l}}
\end{table}

\bibliographystyle{JHEP}
\bibliography{stops}

\providecommand{\href}[2]{#2}\begingroup\raggedright\begin{thebibliography}{10}

\bibitem{Martin:1997ns}
S.~P. Martin, {\it {A Supersymmetry primer}},
  \href{http://xxx.lanl.gov/abs/hep-ph/9709356}{{\tt hep-ph/9709356}}.

\bibitem{Rolbiecki:2013fia}
K.~Rolbiecki and K.~Sakurai, {\it {Light stops emerging in WW cross section
  measurements?}},  {\em JHEP} {\bf 1309} (2013) 004,
  [\href{http://xxx.lanl.gov/abs/1303.5696}{{\tt arXiv:1303.5696}}].

\bibitem{ATOM}
I.-W. Kim, M.~Papucci, K.~Sakurai, and A.~Weiler, {\it {ATOM: Automated Testing
  Of Models}},  \href{http://xxx.lanl.gov/abs/to appear}{{\tt to appear}}.

\bibitem{Drees:2013wra}
M.~Drees, H.~Dreiner, D.~Schmeier, J.~Tattersall, and J.~S. Kim, {\it
  {CheckMATE: Confronting your Favourite New Physics Model with LHC Data}},
  \href{http://xxx.lanl.gov/abs/1312.2591}{{\tt arXiv:1312.2591}}.

\bibitem{ATLAS:2012mec}
{\bf ATLAS} Collaboration, G.~Aad et~al., {\it {Measurement of $W^+W^-$
  production in $pp$ collisions at $\sqrt{s}=7$ TeV with the ATLAS detector and
  limits on anomalous $WWZ$ and $WW\gamma$ couplings}},
  \href{http://xxx.lanl.gov/abs/1210.2979}{{\tt arXiv:1210.2979}}.

\bibitem{Chatrchyan:2013oev}
{\bf CMS} Collaboration, S.~Chatrchyan et~al., {\it {Measurement of $W^+W^-$
  and $ZZ$ production cross sections in $pp$ collisions at $\sqrt{s}=8$ TeV}},
  \href{http://xxx.lanl.gov/abs/1301.4698}{{\tt arXiv:1301.4698}}.

\bibitem{Chatrchyan:2013yaa}
{\bf CMS} Collaboration, S.~Chatrchyan et~al., {\it {Measurement of the
  $W^+W^-$ Cross section in $pp$ Collisions at $\sqrt{s} = 7$ TeV and Limits on
  Anomalous $WW\gamma$ and $WWZ$ couplings}},  {\em Eur.Phys.J.} {\bf C73}
  (2013) 2610, [\href{http://xxx.lanl.gov/abs/1306.1126}{{\tt
  arXiv:1306.1126}}].

\bibitem{Baglio:2013toa}
J.~Baglio, L.~D. Ninh, and M.~M. Weber, {\it {Massive gauge boson pair
  production at the LHC: a next-to-leading order story}},  {\em Phys.Rev.} {\bf
  D88} (2013) 113005, [\href{http://xxx.lanl.gov/abs/1307.4331}{{\tt
  arXiv:1307.4331}}].

\bibitem{Dawson:2013lya}
S.~Dawson, I.~M. Lewis, and M.~Zeng, {\it {Threshold Resummed and Approximate
  NNLO results for $W^+W^-$ Pair Production at the LHC}},  {\em Phys.Rev.} {\bf
  D88} (2013) 054028, [\href{http://xxx.lanl.gov/abs/1307.3249}{{\tt
  arXiv:1307.3249}}].

\bibitem{::2014uva}
{\bf HERAFitter developers' team} Collaboration, P.~Belov et~al., {\it {Parton
  distribution functions at LO, NLO and NNLO with correlated uncertainties
  between orders}},  \href{http://xxx.lanl.gov/abs/1404.4234}{{\tt
  arXiv:1404.4234}}.

\bibitem{Curtin:2012nn}
D.~Curtin, P.~Jaiswal, and P.~Meade, {\it {Charginos Hiding In Plain Sight}},
  {\em Phys.Rev.} {\bf D87} (2013) 031701,
  [\href{http://xxx.lanl.gov/abs/1206.6888}{{\tt arXiv:1206.6888}}].

\bibitem{Curtin:2013gta}
D.~Curtin, P.~Jaiswal, P.~Meade, and P.-J. Tien, {\it {Casting Light on BSM
  Physics with SM Standard Candles}},  {\em JHEP} {\bf 1308} (2013) 068,
  [\href{http://xxx.lanl.gov/abs/1304.7011}{{\tt arXiv:1304.7011}}].

\bibitem{Aad:2011dm}
{\bf ATLAS} Collaboration, G.~Aad et~al., {\it {Measurement of the inclusive
  $W^\pm$ and Z/$\gamma$ cross sections in the electron and muon decay channels
  in $pp$ collisions at $\sqrt{s}=7$ TeV with the ATLAS detector}},  {\em
  Phys.Rev.} {\bf D85} (2012) 072004,
  [\href{http://xxx.lanl.gov/abs/1109.5141}{{\tt arXiv:1109.5141}}].

\bibitem{Chatrchyan:2014mua}
{\bf CMS} Collaboration, S.~Chatrchyan et~al., {\it {Measurement of inclusive W
  and Z boson production cross sections in pp collisions at $\sqrt{s}$ = 8
  TeV}},  {\em Phys.Rev.Lett.} {\bf 112} (2014) 191802,
  [\href{http://xxx.lanl.gov/abs/1402.0923}{{\tt arXiv:1402.0923}}].

\bibitem{Chatrchyan:2013tia}
{\bf CMS} Collaboration, S.~Chatrchyan et~al., {\it {Measurement of the
  differential and double-differential Drell-Yan cross sections in
  proton-proton collisions at $\sqrt{s} =$ 7 TeV}},  {\em JHEP} {\bf 1312}
  (2013) 030, [\href{http://xxx.lanl.gov/abs/1310.7291}{{\tt
  arXiv:1310.7291}}].

\bibitem{Aad:2013iua}
{\bf ATLAS} Collaboration, G.~Aad et~al., {\it {Measurement of the high-mass
  Drell--Yan differential cross-section in pp collisions at $\sqrt{s}=7$ TeV
  with the ATLAS detector}},  {\em Phys.Lett.} {\bf B725} (2013) 223--242,
  [\href{http://xxx.lanl.gov/abs/1305.4192}{{\tt arXiv:1305.4192}}].

\bibitem{Aad:2012twa}
{\bf ATLAS} Collaboration, G.~Aad et~al., {\it {Measurement of $WZ$ production
  in proton-proton collisions at $\sqrt{s}=7$ TeV with the ATLAS detector}},
  {\em Eur.Phys.J.} {\bf C72} (2012) 2173,
  [\href{http://xxx.lanl.gov/abs/1208.1390}{{\tt arXiv:1208.1390}}].

\bibitem{ATLAS:2013fma}
{\bf ATLAS} Collaboration, {\it {A Measurement of WZ Production in
  Proton-Proton Collisions at $\sqrt{s}=8$ TeV with the ATLAS Detector}},
  Tech. Rep. ATLAS-CONF-2013-021, ATLAS-COM-CONF-2013-016, 2013.

\bibitem{CMS:2013qea}
{\bf CMS} Collaboration, {\it {Measurement of WZ production rate}},  Tech. Rep.
  CMS-PAS-SMP-12-006, 2013.

\bibitem{Aad:2012awa}
{\bf ATLAS} Collaboration, G.~Aad et~al., {\it {Measurement of $ZZ$ production
  in $pp$ collisions at $\sqrt{s}=7$ TeV and limits on anomalous $ZZZ$ and
  $ZZ\gamma$ couplings with the ATLAS detector}},  {\em JHEP} {\bf 1303} (2013)
  128, [\href{http://xxx.lanl.gov/abs/1211.6096}{{\tt arXiv:1211.6096}}].

\bibitem{ATLAS:2013gma}
{\bf ATLAS} Collaboration, {\it {Measurement of the total ZZ production cross
  section in proton-proton collisions at $\sqrt{s} = 8$ TeV in 20 fb$^{−1}$
  with the ATLAS detector}},  Tech. Rep. ATLAS-CONF-2013-020,
  ATLAS-COM-CONF-2013-020, 2013.

\bibitem{CMS:2013hea}
{\bf CMS} Collaboration, {\it {Measurement of the ZZ production cross section
  and anomalous trilinear gauge couplings in $\ell\ell\ell'\ell'$ decays at
  $\sqrt{s} = 8$ TeV at the LHC}},  Tech. Rep. CMS-PAS-SMP-13-005, 2013.

\bibitem{ATLAS-CONF-2013-062}
{\bf ATLAS} Collaboration, {\it {Search for squarks and gluinos in events with
  isolated leptons, jets and missing transverse momentum at $\sqrt{s}=8$ TeV
  with the ATLAS detector}},  Tech. Rep. ATLAS-CONF-2013-062, CERN, Geneva,
  Jun, 2013.

\bibitem{Aad:2014nua}
{\bf ATLAS} Collaboration, G.~Aad et~al., {\it {Search for direct production of
  charginos and neutralinos in events with three leptons and missing transverse
  momentum in $\sqrt{s}$ = 8 TeV $pp$ collisions with the ATLAS detector}},
  \href{http://xxx.lanl.gov/abs/1402.7029}{{\tt arXiv:1402.7029}}.

\bibitem{Aad:2014vma}
{\bf ATLAS} Collaboration, G.~Aad et~al., {\it {Search for direct production of
  charginos, neutralinos and sleptons in final states with two leptons and
  missing transverse momentum in pp collisions at $\sqrt{s}$ = 8 TeV with the
  ATLAS detector}},  \href{http://xxx.lanl.gov/abs/1403.5294}{{\tt
  arXiv:1403.5294}}.

\bibitem{ATLAS-CONF-2013-030}
{\bf ATLAS} Collaboration, {\it {Measurements of the properties of the
  Higgs-like boson in the $WW^{(\ast)} \to \ell \nu \ell \nu$ decay channel
  with the ATLAS detector using 25 fb$^{-1}$ of proton-proton collision data}},
   Tech. Rep. ATLAS-CONF-2013-030, CERN, Geneva, Mar, 2013.

\bibitem{ATLAS-CONF-2013-031}
{\bf ATLAS} Collaboration, {\it {Study of the spin properties of the Higgs-like
  particle in the $\boldsymbol{H\to WW^{(\ast)}\to e\nu\mu\nu}$ channel with 21
  fb$^{-1}$ of $\sqrt{s} = 8$ TeV data collected with the ATLAS detector}},
  Tech. Rep. ATLAS-CONF-2013-031, CERN, Geneva, Mar, 2013.

\bibitem{Aad:2014qaa}
{\bf ATLAS} Collaboration, G.~Aad et~al., {\it {Search for direct top-squark
  pair production in final states with two leptons in pp collisions at
  $\sqrt{s}$=8 TeV with the ATLAS detector}},
  \href{http://xxx.lanl.gov/abs/1403.4853}{{\tt arXiv:1403.4853}}.

\bibitem{ATLAS:2013pla}
{\bf ATLAS} Collaboration, {\it {Search for direct top squark pair production
  in final states with one isolated lepton, jets, and missing transverse
  momentum in $\sqrt{s}=8$ TeV $pp$ collisions using 21 fb$^{-1}$ of ATLAS
  data}},  Tech. Rep. ATLAS-CONF-2013-037, ATLAS-COM-CONF-2013-038, CERN,
  Geneva, 2013.

\bibitem{Camargo-Molina:2014pwa}
J.~Camargo-Molina, B.~Garbrecht, B.~O'Leary, W.~Porod, and F.~Staub, {\it
  {Constraining the Natural MSSM through tunneling to color-breaking vacua at
  zero and non-zero temperature}},
  \href{http://xxx.lanl.gov/abs/1405.7376}{{\tt arXiv:1405.7376}}.

\bibitem{Buckley:2010ar}
A.~Buckley, J.~Butterworth, L.~Lonnblad, H.~Hoeth, J.~Monk, et~al., {\it {Rivet
  user manual}},  \href{http://xxx.lanl.gov/abs/1003.0694}{{\tt
  arXiv:1003.0694}}.

\bibitem{deFavereau:2013fsa}
J.~de~Favereau, C.~Delaere, P.~Demin, A.~Giammanco, V.~Lemaître, et~al., {\it
  {DELPHES 3, A modular framework for fast simulation of a generic collider
  experiment}},  \href{http://xxx.lanl.gov/abs/1307.6346}{{\tt
  arXiv:1307.6346}}.

\bibitem{Cacciari:2011ma}
M.~Cacciari, G.~P. Salam, and G.~Soyez, {\it {FastJet User Manual}},  {\em
  Eur.Phys.J.} {\bf C72} (2012) 1896,
  [\href{http://xxx.lanl.gov/abs/1111.6097}{{\tt arXiv:1111.6097}}].

\bibitem{Cacciari:2005hq}
M.~Cacciari and G.~P. Salam, {\it {Dispelling the $N^{3}$ myth for the $k_t$
  jet-finder}},  {\em Phys.Lett.} {\bf B641} (2006) 57--61,
  [\href{http://xxx.lanl.gov/abs/hep-ph/0512210}{{\tt hep-ph/0512210}}].

\bibitem{Cacciari:2008gp}
M.~Cacciari, G.~P. Salam, and G.~Soyez, {\it {The Anti-k(t) jet clustering
  algorithm}},  {\em JHEP} {\bf 0804} (2008) 063,
  [\href{http://xxx.lanl.gov/abs/0802.1189}{{\tt arXiv:0802.1189}}].

\bibitem{Lester:1999tx}
C.~Lester and D.~Summers, {\it {Measuring masses of semiinvisibly decaying
  particles pair produced at hadron colliders}},  {\em Phys.Lett.} {\bf B463}
  (1999) 99--103, [\href{http://xxx.lanl.gov/abs/hep-ph/9906349}{{\tt
  hep-ph/9906349}}].

\bibitem{Barr:2003rg}
A.~Barr, C.~Lester, and P.~Stephens, {\it {m(T2): The Truth behind the
  glamour}},  {\em J.Phys.} {\bf G29} (2003) 2343--2363,
  [\href{http://xxx.lanl.gov/abs/hep-ph/0304226}{{\tt hep-ph/0304226}}].

\bibitem{Cheng:2008hk}
H.-C. Cheng and Z.~Han, {\it {Minimal Kinematic Constraints and m(T2)}},  {\em
  JHEP} {\bf 0812} (2008) 063, [\href{http://xxx.lanl.gov/abs/0810.5178}{{\tt
  arXiv:0810.5178}}].

\bibitem{Bahr:2008pv}
M.~Bahr, S.~Gieseke, M.~Gigg, D.~Grellscheid, K.~Hamilton, et~al., {\it
  {Herwig++ Physics and Manual}},  {\em Eur.Phys.J.} {\bf C58} (2008) 639--707,
  [\href{http://xxx.lanl.gov/abs/0803.0883}{{\tt arXiv:0803.0883}}].

\bibitem{Bellm:2013lba}
J.~Bellm, S.~Gieseke, D.~Grellscheid, A.~Papaefstathiou, S.~Platzer, et~al.,
  {\it {Herwig++ 2.7 Release Note}},
  \href{http://xxx.lanl.gov/abs/1310.6877}{{\tt arXiv:1310.6877}}.

\bibitem{Beenakker:1996ch}
W.~Beenakker, R.~Hopker, M.~Spira, and P.~Zerwas, {\it {Squark and gluino
  production at hadron colliders}},  {\em Nucl.Phys.} {\bf B492} (1997)
  51--103, [\href{http://xxx.lanl.gov/abs/hep-ph/9610490}{{\tt
  hep-ph/9610490}}].

\bibitem{Beenakker:1997ut}
W.~Beenakker, M.~Kramer, T.~Plehn, M.~Spira, and P.~Zerwas, {\it {Stop
  production at hadron colliders}},  {\em Nucl.Phys.} {\bf B515} (1998) 3--14,
  [\href{http://xxx.lanl.gov/abs/hep-ph/9710451}{{\tt hep-ph/9710451}}].

\bibitem{Beenakker:2009ha}
W.~Beenakker, S.~Brensing, M.~Kramer, A.~Kulesza, E.~Laenen, et~al., {\it
  {Soft-gluon resummation for squark and gluino hadroproduction}},  {\em JHEP}
  {\bf 0912} (2009) 041, [\href{http://xxx.lanl.gov/abs/0909.4418}{{\tt
  arXiv:0909.4418}}].

\bibitem{Beenakker:2010nq}
W.~Beenakker, S.~Brensing, M.~Kramer, A.~Kulesza, E.~Laenen, et~al., {\it
  {Supersymmetric top and bottom squark production at hadron colliders}},  {\em
  JHEP} {\bf 1008} (2010) 098, [\href{http://xxx.lanl.gov/abs/1006.4771}{{\tt
  arXiv:1006.4771}}].

\bibitem{Beenakker:1999xh}
W.~Beenakker, M.~Klasen, M.~Kramer, T.~Plehn, M.~Spira, et~al., {\it {The
  Production of charginos / neutralinos and sleptons at hadron colliders}},
  {\em Phys.Rev.Lett.} {\bf 83} (1999) 3780--3783,
  [\href{http://xxx.lanl.gov/abs/hep-ph/9906298}{{\tt hep-ph/9906298}}].

\bibitem{Nadolsky:2008zw}
P.~M. Nadolsky, H.-L. Lai, Q.-H. Cao, J.~Huston, J.~Pumplin, et~al., {\it
  {Implications of CTEQ global analysis for collider observables}},  {\em
  Phys.Rev.} {\bf D78} (2008) 013004,
  [\href{http://xxx.lanl.gov/abs/0802.0007}{{\tt arXiv:0802.0007}}].

\bibitem{Dreiner:2012sh}
H.~Dreiner, M.~Krämer, and J.~Tattersall, {\it {Exploring QCD uncertainties
  when setting limits on compressed supersymmetric spectra}},  {\em Phys.Rev.}
  {\bf D87} (2013), no.~3 035006,
  [\href{http://xxx.lanl.gov/abs/1211.4981}{{\tt arXiv:1211.4981}}].

\bibitem{Dreiner:2012gx}
H.~K. Dreiner, M.~Kramer, and J.~Tattersall, {\it {How low can SUSY go?
  Matching, monojets and compressed spectra}},  {\em Europhys.Lett.} {\bf 99}
  (2012) 61001, [\href{http://xxx.lanl.gov/abs/1207.1613}{{\tt
  arXiv:1207.1613}}].

\bibitem{Khachatryan:2014qwa}
{\bf CMS} Collaboration, V.~Khachatryan et~al., {\it {Searches for electroweak
  production of charginos, neutralinos, and sleptons decaying to leptons and W,
  Z, and Higgs bosons in pp collisions at 8 TeV}},
  \href{http://xxx.lanl.gov/abs/1405.7570}{{\tt arXiv:1405.7570}}.

\bibitem{Chatrchyan:2013xna}
{\bf CMS} Collaboration, S.~Chatrchyan et~al., {\it {Search for top-squark pair
  production in the single-lepton final state in pp collisions at $\sqrt{s}$ =
  8 TeV}},  {\em Eur.Phys.J.} {\bf C73} (2013) 2677,
  [\href{http://xxx.lanl.gov/abs/1308.1586}{{\tt arXiv:1308.1586}}].

\bibitem{Chatrchyan:2012paa}
{\bf CMS} Collaboration, S.~Chatrchyan et~al., {\it {Search for new physics in
  events with same-sign dileptons and $b$ jets in $pp$ collisions at
  $\sqrt{s}=8$ TeV}},  {\em JHEP} {\bf 1303} (2013) 037,
  [\href{http://xxx.lanl.gov/abs/1212.6194}{{\tt arXiv:1212.6194}}].

\bibitem{CMS:2013dea}
{\bf CMS} Collaboration, {\it {Search for electroweak production of charginos,
  neutralinos, and sleptons using leptonic final states in pp collisions at 8
  TeV}},  Tech. Rep. CMS-PAS-SUS-13-006, 2013.

\bibitem{ATLAS-CONF-2013-089}
{\bf ATLAS} Collaboration, {\it {Search for strongly produced supersymmetric
  particles in decays with two leptons at $\sqrt{s}$ = 8 TeV}},  Tech. Rep.
  ATLAS-CONF-2013-089, CERN, Geneva, Aug, 2013.

\bibitem{Stats}
J.~Conway, {\it {Calculation of Cross Section Upper Limits Combining Channels
  Incorporating Correlated and Uncorrelated Systematic Uncertainties}},
  \href{http://xxx.lanl.gov/abs/CDF/PUB/STATISTICS/PUBLIC/6428}{{\tt
  CDF/PUB/STATISTICS/PUBLIC/6428}}.

\bibitem{Barr:2005dz}
A.~J. Barr, {\it {Measuring slepton spin at the LHC}},  {\em JHEP} {\bf 02}
  (2006) 042, [\href{http://xxx.lanl.gov/abs/hep-ph/0511115}{{\tt
  hep-ph/0511115}}].

\bibitem{Diener:2009ee}
R.~Diener, S.~Godfrey, and T.~A. Martin, {\it {Using Final State
  Pseudorapidities to Improve s-channel Resonance Observables at the LHC}},
  {\em Phys.Rev.} {\bf D80} (2009) 075014,
  [\href{http://xxx.lanl.gov/abs/0909.2022}{{\tt arXiv:0909.2022}}].

\bibitem{MoortgatPick:2011ix}
G.~Moortgat-Pick, K.~Rolbiecki, and J.~Tattersall, {\it {Early spin
  determination at the LHC?}},  {\em Phys.Lett.} {\bf B699} (2011) 158--163,
  [\href{http://xxx.lanl.gov/abs/1102.0293}{{\tt arXiv:1102.0293}}].

\bibitem{Buckley:2014fqa}
M.~R. Buckley, T.~Plehn, and M.~J. Ramsey-Musolf, {\it {Stop on Top}},
  \href{http://xxx.lanl.gov/abs/1403.2726}{{\tt arXiv:1403.2726}}.

\bibitem{Drees:1990dx}
M.~Drees and K.~Hagiwara, {\it {Supersymmetric Contribution to the Electroweak
  $\rho$ Parameter}},  {\em Phys.Rev.} {\bf D42} (1990) 1709--1725.

\bibitem{Heinemeyer:2013dia}
S.~Heinemeyer, W.~Hollik, G.~Weiglein, and L.~Zeune, {\it {Implications of LHC
  search results on the W boson mass prediction in the MSSM}},  {\em JHEP} {\bf
  1312} (2013) 084, [\href{http://xxx.lanl.gov/abs/1311.1663}{{\tt
  arXiv:1311.1663}}].

\bibitem{ArkaniHamed:2006mb}
N.~Arkani-Hamed, A.~Delgado, and G.~Giudice, {\it {The Well-tempered
  neutralino}},  {\em Nucl.Phys.} {\bf B741} (2006) 108--130,
  [\href{http://xxx.lanl.gov/abs/hep-ph/0601041}{{\tt hep-ph/0601041}}].

\bibitem{Barger:2005hb}
V.~Barger, P.~Langacker, and H.-S. Lee, {\it {Lightest neutralino in extensions
  of the MSSM}},  {\em Phys.Lett.} {\bf B630} (2005) 85--99,
  [\href{http://xxx.lanl.gov/abs/hep-ph/0508027}{{\tt hep-ph/0508027}}].

\bibitem{Kraml:2008zr}
S.~Kraml, A.~Raklev, and M.~White, {\it {NMSSM in disguise: Discovering
  singlino dark matter with soft leptons at the LHC}},  {\em Phys.Lett.} {\bf
  B672} (2009) 361--366, [\href{http://xxx.lanl.gov/abs/0811.0011}{{\tt
  arXiv:0811.0011}}].

\bibitem{Kim:2014noa}
J.~S. Kim and T.~S. Ray, {\it {The Higgsino-Singlino World}},
  \href{http://xxx.lanl.gov/abs/1405.3700}{{\tt arXiv:1405.3700}}.

\bibitem{Curtin:2014zua}
D.~Curtin, P.~Meade, and P.-J. Tien, {\it {Natural SUSY in Plain Sight}},
  \href{http://xxx.lanl.gov/abs/1406.0848}{{\tt arXiv:1406.0848}}.

\end{thebibliography}\endgroup

\end{document}